\documentstyle[epsfig,longtable]{aipproc}

\begin{document}
\title{A Comparison of the LSND and\\ KARMEN $\bar\nu$ Oscillation
Experiments}

\author{Steven J. Yellin}
\address{Univ. of Calif. at Santa Barbara, yellin@slac.stanford.edu}

\maketitle

\begin{abstract}
The LSND and KARMEN $\nu$ oscillation experiments are compared in order to
clarify the significance of the apparent disagreement between their results.
\end{abstract}

\subsection*{Introduction}

\begin{figure}[t!] 
\centerline{\epsfig{file=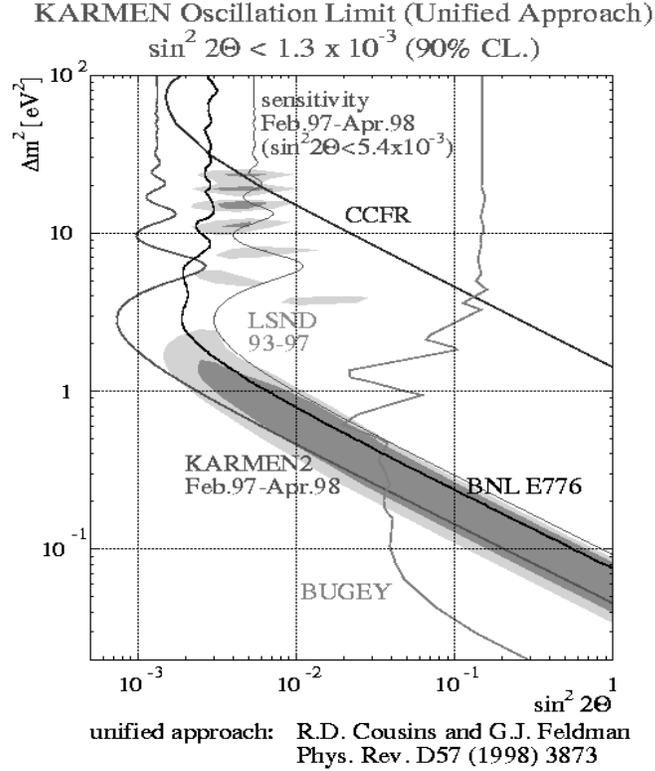,height=5in,width=4in}}
\vspace{10pt}
\caption{Comparison of preliminary LSND $\bar\nu_\mu\rightarrow\bar\nu_e$
results with those of KARMEN and other experiments.  Everything to the right
of the leftmost curve is excluded to 90\% confidence level by KARMEN; the
light and dark filled regions are ones favored by LSND.} 
\label{fig1}
\end{figure}

The LSND experiment, performed at the Los Alamos National Laboratory, and
the KARMEN experiment, performed at the ISIS spallation facility at
Rutherford Appleton Laboratory, give results that appear to be in conflict.
Both groups experimentally investigated the possibility of
$\bar\nu_\mu\rightarrow \bar\nu_e$.  They expressed their results in terms
of a two-neutrino oscillation model in which only the
muon and electron neutrinos are relevant, so that
the probability of observing an electron neutrino of energy $E_\nu$ MeV
a distance $L$ meters away from formation of a muon neutrino is
\begin{equation}
P=sin^22\theta\,sin^2\frac{1.27\delta m^2\,L}{E_\nu},
\end{equation}
with $\theta$ the mixing angle and $\delta m^2$ the difference in $eV^2$
between the squared masses.  LSND has published evidence
for a signal\cite{LSNDdar}, and includes more recent data in figure 
 \ref{fig1}.  But recent KARMEN results, presented in conferences such as
Neutrino '98\cite{Neut98} and shown
in the same figure, exclude almost all of the favored LSND region.

Both experiments have competent personnel (this author is in LSND), and both
have been working a long enough time to eliminate serious mistakes.
How, then, do the experiments differ, and are their results
really inconsistent?

\subsection*{Comparison of Experimental Methods}

Both experiments produce neutrinos with intense, 800 MeV proton beams.
These beams hit targets and create, among other things, $\pi^+$ mesons,
which decay mainly into $\mu^+\nu_\mu$, and the $\mu^+$ mesons almost all
decay at rest into $e^+\nu_e\bar\nu_\mu$.  The experiments investigate
whether the resulting $\bar\nu_\mu$ oscillate into $\bar\nu_e$, which are
detected by the reaction $\bar\nu_e p\rightarrow e^+n$.  Both the $e^+$
and the $n$ are detected in a large, liquid scintillator detector,
the $n$ by delayed scintillation light caused by the $\gamma$ from neutron
capture in the detector.  One might think that $\pi^-$ would be produced
in the target along with $\pi^+$, and that the antiparticle decay chain
would lead to serious $\bar\nu_e$ background from $\mu^-\rightarrow
e^-\bar\nu_e\nu_\mu$.  But the beam and target consist of matter, rather
than antimatter, and that makes a great difference between the $\pi^+$ and
$\pi^-$ decay chains.  Production of $\pi^-$ is about 1/8 of $\pi^+$
production, $\pi^-$ decay is suppressed by a factor of 20 because these
negative particles are attracted to positive nuclei in the target and are
usually absorbed before they can decay, and $\mu^-$ decay is suppressed by
a factor of 10 for the same reason.  Thus $\bar\nu_e$ production in the
target is more than three orders of magnitude below $\bar\nu_\mu$ production,
allowing experiments to be sensitive to very small probability of
$\bar\nu_\mu\rightarrow\bar\nu_e$ oscillation.

There are important differences between the beams of the two
experiments.  The LSND target was $H_2O$ followed by a drift
space, allowing $\pi^+\rightarrow \mu^+\nu_\mu$ decay in flight at
the 3.4\% level, while KARMEN's target was $Ta+D_2O$ with no
drift space; so only 0.1\% of its $\pi^+$ mesons decayed in flight.  Thus
LSND could, and did\cite{LSNDDIF}, also measure the charge conjugate oscillation,
$\nu_\mu\rightarrow\nu_e$, by detecting higher energy decay in flight
neutrinos with $\nu_e C\rightarrow e^- X$.  Other significant differences
were that LSND had 1 ma proton beam intensity instead of KARMEN's 0.2 ma,
and LSND had a much larger duty
factor than KARMEN, 0.07 instead of $5\times 10^{-4}$.  KARMEN's tiny duty
factor was a great advantage, because the two experiments suppressed and
measured cosmic background by comparing data with and without beam.

There are also important differences between the two detectors.
The LSND detector was centered 30 meters from the target, rather than the
17.6 meters of KARMEN.  This greater distance of LSND lowered
the neutrino flux, but increased the oscillation probability with low
$\delta m^2$,
and made the experiment sensitive to somewhat different $\delta m^2$.
The LSND detector was over three times as large as KARMEN's, 167 tons of
scintillator instead of 56 tons.  LSND had a single tank
whose walls were covered with photomultiplier tubes.  KARMEN, on the other
hand, was segmented into 512 modules.  This segmentation gave KARMEN
better position resolution, especially for the low energy $\gamma$
produced upon capture of the neutron from $\bar\nu_e p\rightarrow e^+n$.
LSND used a much lower concentration of scintillator in the liquid.
KARMEN therefore had a factor of four better energy resolution, which can
be useful for distinguishing between neutron capture $\gamma$'s and the lower
energy accidental $\gamma$'s from radioactivity.  But KARMEN could not
distinguish between neutrons and $e^+$ nearly as well as LSND.  Because the
LSND detector was one big tank with a low concentration of scintillator, the
requirement of Cherenkov ring detection could
exclude almost all high energy neutrons produced by cosmic muons, and the
$e^+$ direction could be measured.  Two
other advantages of LSND were a) surrounding the detector was a superior
veto against cosmic muons, and b) it had more shielding against neutrons
from the target.  The first of these two advantages was eliminated in 1996 by
an upgrade of KARMEN's veto.  The KARMEN data shown in figure \ref{fig1} are
based entirely on 2897 coulombs of integrated proton beam taken
after their upgrade over a three month period.
LSND's results shown in that figure are based on over 20000
coulombs collected between 1993 and 1997.

\subsection*{Comparison of the Analyses}

LSND's analysis was more complex than what KARMEN used for its
Neutrino '98\cite{Neut98} presentation.  LSND distinguished $e^+$ from
high energy cosmogenic neutrons by requiring
the showers to produce a good Cherenkov ring, to have
a larger portion of fast light than would come from pure scintillation, and
to be confined in position.  The $e^+$ energy was
required to be high enough to avoid $\nu_e\,^{12}C\rightarrow e^-n\,^{11}N$
contamination, but not too high to be from oscillation of $\bar\nu_\mu$ from
decay at rest of $\mu^+$: $20\ MeV < E_e < 60\ MeV$.  KARMEN's energy cut was
$20\ MeV < E_e < 50\ MeV$.  Both LSND and KARMEN
suppressed cosmogenic neutron background by excluding events with a veto
counter signal.  LSND also excluded events with evidence of something other
than a neutrino entering the detector by looking for even small veto signals
and by using the $e^+$ direction and position to tell if it entered from
outside.  LSND used information from both before and after the $e^+$
to exclude events caused by $\mu\rightarrow e\nu\nu$, and KARMEN used only
information from before the $e^+$ for that purpose.

Once events with an apparent $e^+$ were selected, both KARMEN and LSND
selected events with a delayed $\gamma$ from neutron capture.
Both experiments required the $\gamma$ to be near in both space and time
to the $e^+$ and to have a reasonable energy.  But KARMEN made its selection
through cuts in position, time, and energy, while LSND needed a more
elaborate analysis since it had poorer resolution in these parameters.  For
LSND, the product of the position, time, and energy
distributions for correlated neutron capture events was divided by the
similar product for accidental $\gamma$'s to form a likelihood ratio, $R$.
High $R$ corresponds to a $\gamma$ that appears more like a neutron capture
correlated with the $e^+$ production, and low $R$ corresponds to a $\gamma$
that appears to be accidental.  LSND could then either cut on high R in order
to select clean candidates, or it could fit the $R$ distribution to measure
and subtract the contribution from accidental $\gamma$'s.  It did both.

KARMEN used plausible criteria to optimize cuts by computer, while
some of LSND's cuts were chosen with a less systematic attempt to avoid
human biases.  Instead, LSND verified that its results were not very
sensitive to variations of cuts.

In order to exclude 
$\nu_e C\rightarrow e^- X$, which almost always has measured $e^-$ energy
below 36 MeV, and which is the largest background with no correlated
$\gamma$, LSND cleaned its sample by further requiring $E_e>36$ MeV.
It then required $R>30$
(nearly unambiguous correlated $\gamma$), leaving 22 beam-on events
and 36 beam-off events. Since LSND (beam-on time)/(beam-off time)=0.070,
cosmic beam-on background is estimated to be
$0.070 \times 36 = 2.5$ events.  There was a beam-associated background
of 0.4 events with a correlated photon from $\bar\nu_e$ produced in the beam
dump, and the measured $R$ distribution of accidentals was used to estimate
1.7 events with an accidental $\gamma$.  This left LSND with a
measured excess of 17.4 events, a signal for which conventional processes
cannot account.  The probability that this was due to a statistical
fluctuation is $4.1\times 10^{-8}$.

KARMEN, after its cuts, found zero beam-on candidates with an expected
background of $2.88\pm 0.13$.  The frequentist ``unified approach'' of
Feldman and Cousins\cite{Feldman}, led to the 90\% confidence level
exclusion limit shown in figure \ref{fig1}.

LSND's contribution to figure \ref{fig1} was done without the tighter
$E_e$ and $R$ cuts used to establish the existence of apparent
oscillation events.  Instead the $R$ distribution was used to measure the
background from accidental $\gamma$'s, and that background was used in a
likelihood function of $\sin^2 2\theta$ and $\delta m^2$, based on the
data's distribution in $R$, $e^+$ position, direction, and energy.  The
largest likelihood regions are the favored ones shown in the figure.

\subsection*{Critique}

KARMEN's 90\% confidence region is misleading.  A 90\% frequentist confidence
region is one chosen by a {\bf method} expected to give the truth 90\% of the
time it's used.  That doesn't mean that {\bf this time} the method has a
90\% probability of giving the truth.  For KARMEN's particular experimental
result, frequentist methods can give a region with much less than 90\%
probability of giving the truth.  For example, if KARMEN had used
the conventional frequentist confidence region, instead of the
``unified approach'', it would have zero percent probability of giving
the truth, for the region would be empty, excluding both $sin^22\theta>0$
and $sin^22\theta=0$!  This happens because KARMEN found no events, a
somewhat surprisingly low result even under the assumption that
$sin^2 2\theta=0$.  While the ``unified approach'' cannot give an empty
region, it can give one that is too small.  In the absence of a priori
assumptions one cannot prove that a non-empty region is too small.  But if
one takes the usual Bayesian a priori uniform distribution in the unknown
parameters ($sin^2 2\theta$ and $\delta m^2$) then the region allowed by
KARMEN with the ``unified approach'' covers less than half of that allowed
by the 90\% Bayesian probability.

\begin{figure}[t!] 
\centerline{\epsfig{file=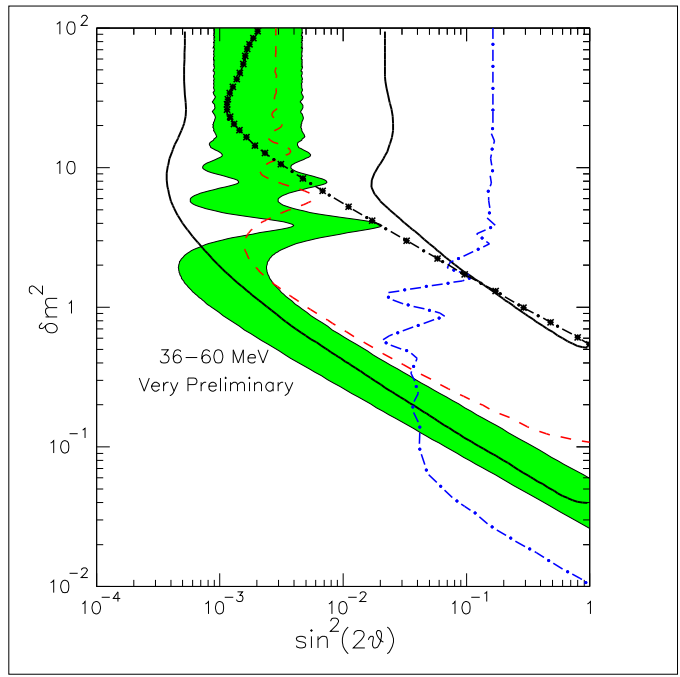,height=3.5in,width=3.5in}}
\vspace{10pt}
\caption{Comparison of preliminary LSND $\bar\nu_\mu\rightarrow\bar\nu_e$
results (filled region) with those of KARMEN (dashed curve), Bugey
(dot-dashed), and NOMAD (dot-x-dash).
KARMEN's exclusion curve is 90\% Bayesian confidence level,
while the LSND region is bounded by 90\% Bayesian upper and lower limits with
$36<E_e<60$ MeV.  The pair of smooth curves surrounding
the LSND region gives the LSND 90\% confidence region for
$\nu_\mu\rightarrow \nu_e$ oscillation.}
\label{fig2}
\end{figure}

Figure \ref{fig1} shows LSND regions surrounded by contours of 0.1 and 0.01
in (likelihood)/(maximum likelihood).  These favored regions can be
misleading if interpreted as confidence regions.  I therefore
compared KARMEN's Bayesian 90\% limit with LSND's data reanalyzed to give
90\% Bayesian upper and lower limits.  For simplicity,
the fit ignored position and energy information.  But unlike the figure
$\ref{fig1}$ result, this fit itself used $R$ to correct for
accidentals, instead of getting the correction elsewhere.  Figure \ref{fig2}
shows the result, based on the same events as
figure \ref{fig1}, but with the effect of uncertain backgrounds lessened by
requiring $E_e>36$ MeV.  Not only is there no inconsistency between LSND and
KARMEN, but the figure also shows that LSND's $\nu_\mu\rightarrow \nu_e$ 90\%
frequentist confidence region\cite{LSNDDIF} is consistent with this way of
analyzing LSND's $\bar\nu_\mu\rightarrow\bar\nu_e$ data.

The LSND analysis of figure \ref{fig2} can be replaced by one whose
requirement of $E_e>36$ MeV is replaced by $E_e>20$ MeV.  In that case,
the lower LSND 90\% limit shifts up
almost a factor of four relative to that of figure \ref{fig2}.
This probably happens because more high $R$ events are seen than
are expected from intermediate $R$.  Since low $E_e$ data have especially high
background, high $R$, clean, events contribute more strongly to a fit with
more low energy contamination.  Fit results might then be increased
by an upward statistical fluctuation in the number of high $R$ events.

The alternative analyses shown in figure \ref{fig2} are in some ways inferior
to the ones LSND and KARMEN have reported elsewhere.
It is better to include $E_e$ and event position in the LSND
analysis.  Frequentist methods like the one KARMEN used do depend less
on controversial a priori assumptions of experimentalists.  The point of figure
{\ref{fig2} is that statistical fluctuations, combined with the
details of the method of experimental analysis, still allow too much freedom
in the experimental results for us to conclude that the two experiments
really disagree.

LSND at the time of this conference is just finishing up with the last
of its data collection, and will not increase its statistics much over
what has already been reported.  But LSND analysis is still underway.  For the
$\nu_\mu \rightarrow \nu_e$ analysis\cite{LSNDDIF}, LSND significantly
improved particle identification, and position and energy resolution, all
while using methods of cut selection that avoided human bias better than
past $\bar\nu_\mu \rightarrow\bar\nu_e$ analyses.  The
$\nu_\mu \rightarrow \nu_e$ results were also displayed as a true frequentist
confidence region, unlike past LSND $\bar\nu_\mu \rightarrow\bar\nu_e$
results.  What was learned and
invented for $\nu_\mu \rightarrow \nu_e$ is being applied to a joint analysis
of both $\nu_\mu$ and $\bar\nu_\mu$ data.

The upgraded KARMEN will eventually have much more data than were reported
at Neutrino '98, and its analysis is also improving.

Perhaps the two experiments will eventually resolve their differences.

\end{document}